# Two-Dimensional Ferroelectric Photonic Crystals: Optics and Band Structure

Sevket SIMSEK, Amirullah M.MAMEDOV and Ekmel OZBAY

Bilkent University, Nanotechnology Research Center (NANOTAM), Ankara, Turkey

**Abstract**

In this report we present an investigation of the optical properties and band structure calculations for the photonic structures based on the functional materials- ferroelectrics. A theoretical approach to the optical properties of the 2D and 3D photonic crystals which yields further insight in the phenomenon of the reflection from different families of lattice planes in relation to the presence of photonic gaps or photonic bands. We calculate the photonic bands and optical properties of $LiNbO_3$ based photonic crystals. Calculations of reflection and transmission spectra show the features correspond to the onset of diffraction, as well as to additional reflectance structures at large values of the angle of incidence.

## I. Introduction

Over the past ten years, there has been great interest in photonic crystal lattices (PCLs) due to their ability to manipulate light and great potential for technological applications in nanophotonics and integrated optoelectronic devices[1-3]. PCLs are artificial materials; optical and electronic properties of two (2D)- and three (3D)-dimensional PCLs are intensively studied with the goal of achieving control of electromagnetic propogation, and, especially, a complete photonic band gap in all directions.[2]

The properties of PCLs strongly depend on the configuration of the selected materials, which cannot be modified after fabrication. Also, photonic band gap (PBG) structure depends on the refractive indices of the selected materials. If the PBG structure of the PCLs can be externally modulated by some other means, the PCLs can be applicable as active elements of many optoelectronic devices. Therefore, as photonic crystalline material we used $LiNbO_3$ ferroelectric material, that has interesting properties for the implementation of active PCLs.

In this paper, we study the electronic band structure and optical properties of the 2D and 3D $LiNbO_3$ based PCLs with square lattices by using FDTD technique, that based on computing eigenstates of Maxwell's equations for periodical dielectric systems [4].

## II. METHOD OF CALCULATION

In our simulation, we employed the finite-difference time domain (FDTD) technique which implies the solution of Maxwell equations with centered finite-difference expressions for the space and time derivatives. FDTD schemes are especially promising for the investigation of PBG structures, as they provide an opportunity of analyzing the spatial distribution of the electromagnetic field in PBG structure. All calculations in this work was performed using OptiFDTD package [5]. OptiFDTD enables to design, analyze and test modern passive and nonlinear photonic components for wave propagation, scattering, reflection, diffraction, polarization and nonlinear phenomena. The core program of OptiFDTD is based on the Finite-Difference Time-Domain (FDTD) algorithm with second-order numerical accuracy and the most advanced boundary conditions - Uniaxial Perfectly Matched Layer (UPML).

The algorithm solves both electric and magnetic fields in temporal and spatial domain using the full-vector differential form of Maxwell's coupled curl equations. This allows for arbitrary model geometries and places no restriction on the material properties of the devices.

The FDTD is characterized by the solution of Maxwell's curl equations in the time domain after replacement of the derivatives in them by finite differences. Since it is a time-domain method, FDTD solutions can cover a wide frequency range with a single simulation run, and treat nonlinear material properties in a natural way. The time-dependent Maxwell's equations (in partial differential form) are discretized using central-difference approximations to the space and time partial derivatives. The resulting finite-difference equations are solved in either software or hardware in a leapfrog manner: the electric field vector components in a volume of space are solved at a given instant in time; then the magnetic field vector components in the same spatial volume are solved at the next instant in time; and the process is repeated over and over again until the desired transient or steady-state electromagnetic field behavior is fully evolved [4].

It is well known that $LiNbO_3$ is an important material because of showing nonlinear behavior. Here, a rectangle lattice with various types scatterer in a dielectric background has been considered at the effect of various parameters on a complete band gap has been investigated. The permittivities of air and $LiNbO_3$ are taken 1 and 5, respectively. We use Lorentz_Drude model for the permittivity of Al and we consider first five resonance terms [6]. Resonance frequency, plasma frequency, strength and damping frequency for Al is given Table 1. The $LiNbO_3$ cylinders and spheres are 0.04 μm in diameter and cylinders are 0.06 μm in height. They are arranged in square mesh with a lattice constant of a=0.1μm.

### III. Result And Discussions

#### A. Photonic Band Gap Calculation

The goal of photonic band structure calculation is the solution of the wave equation for the PCLs i.e., for a strictly periodic array of ferroelectric material. The resulting dispersion relation and associated mode structure may then be further processed to derive some physical quantities such as optical properties and imagine map of electric field in x- and y- direciton.

In Figure 1-2 presents us the unit cell and associated Brillouin zone for $LiNbO_3$ based PCLs, geometry of calculated structures in PCLs. In Figure 3, we show the hybrid band structures for TE- and TM polarized radiations in 2D PCLs consisting of a square lattice (lattice constant a) of the spheres and cylindirical dielectric rods. These strcutures exhibit three complete bandgaps in the X-M and M-Γ high symmetry directions.

### B. Transmission and Reflection in Ferroelectric based PCL

Let us consider a photonic crystal structure in the form described in Fig (1-2). This structure is characterized by the dielectric permittivities of the layers of the medium ε. The thicknesses of the layers are $d_1$ and $d_2$ and the structure period is $a=d_1+d_2$. In the Lorentz-Drude approximation we study the transmission and reflection spectra of polarized waves propagating in the xz-plane. For the structure unders study, the electric field distribution in the layers has the form

$$E_X(n,t) = \left[ A_n e^{i\alpha_n(z-z_n)} + B_n e^{i\alpha_n(z-z_n)} \right] e^{-i\omega t}$$

Where $A_n$ and $B_n$ are the amplitudes of the incident and reflected waves in the n.th layer, $\alpha_n = (\omega/c)\sqrt{\varepsilon_n - \sin^2\theta}$, and θ is the angle of incidence of the radiation. The magnetic field distribution in the layers has the same classical expresseion.

From the continuity of the electric and magnetic fields at the interface $z_n=z_{n+1}$ between layers, we can obtain the system of equations which can be represented as the matrix equation. Let us now study particular features of the transmission and reflection spectra of PCL by using FDTD software.

Figure 4 shows a typical seed band structure of the transmission spectrum of PCL with $\varepsilon_1= 1.0$ (air), $\varepsilon_2= 5$ ($LiNbO_3$) in the case of normal incidence of radiation on the structure. The first band gap of this PCL (rods) lies in the wavelength region between 125.6 nm and 135.6 nm and the frequency $\omega_0$ lies in the center of this region (130 nm), the next band gap lies in the wavelength between 180 nm and 199.4 nm and the frequency $\omega_0$ lies in the center of this region (189.7 nm) and the last band gap lies in the wavelength between 269.7 nm and 303.1 nm and the frequency $\omega_0$ lies in the center of this region (286.4 nm). For the PCL (sphere) the first band gap of this PCL (rods) lies in the wavelength region between 116.4 nm and 128.1 nm and the frequency $\omega_0$ lies in the center of this region (122.3 nm), the second band gap lies in the wavelength between 159.1 nm and 181.6 nm and the frequency $\omega_0$ lies in the center of this region (170.4 nm) and the last band gap lies in the wavelength between 245.3 nm and 268.7 nm and the frequency $\omega_0$ lies in the center of this region (257 nm).

## IV. Conclusion

In this paper a study of the photonic band structure for $LiNbO_3$ based artificial PCL'c with square primitive lattice is reported. The reduce photonic bands and reflectance and transmittance spectra of $LiNbO_3$ based PCLs with two different geometrical configuration (rods and spheres) are calculated by using FDTD scheme.

This paper were presented at 21 st International Symposium on Applications of Ferroelectrics (ISAF ECAPD PFM 2012 Averio, Portugal) and accepted for publication in Ferroelectrics (2013).

**Table and Figure legends**

Table 1. Strengths, plasma frequencies, resonant frequencies and damping frequencies for Al substrate

Figure 1. Unit cell and associated Brillouin zone for square lattice.

Figure 2. Geometry of the calculation for $LiNbO_3$ based structures on Al substrate: cylinders (a) and spheres (b)

Figure 3. Hybrid photonic band structure of $LiNbO_3$ based PCLs on Al substrate: cylinders (a) and spheres (b)

Figure 4. Transmitted power spectrum of $LiNbO_3$ PCLs on Al substrate: cylinders (a) and spheres (b)

**Table 1.**

| Resonance terms | Strength | Plasma Frequency(rad/s) | Resonant Frequency(rad/s) | Damping Frequency(rad/s) |
|---|---|---|---|---|
| 1 | $5.23 \times 10^{-1}$ | $2.2758 \times 10^{16}$ | 0 | $7.14 \times 10^{13}$ |
| 2 | $2.27 \times 10^{-1}$ | $2.2758 \times 10^{16}$ | $2.461 \times 10^{14}$ | $5.059 \times 10^{14}$ |
| 3 | $5.23 \times 10^{-2}$ | $2.2758 \times 10^{16}$ | $2.346 \times 10^{15}$ | $4.74 \times 10^{14}$ |
| 4 | $1.66 \times 10^{-1}$ | $2.2758 \times 10^{16}$ | $2.747 \times 10^{15}$ | $2.053 \times 10^{15}$ |
| 5 | $3.0 \times 10^{-2}$ | $2.2758 \times 10^{16}$ | $5.276 \times 10^{15}$ | $5.138 \times 10^{15}$ |

**Figure 1.**

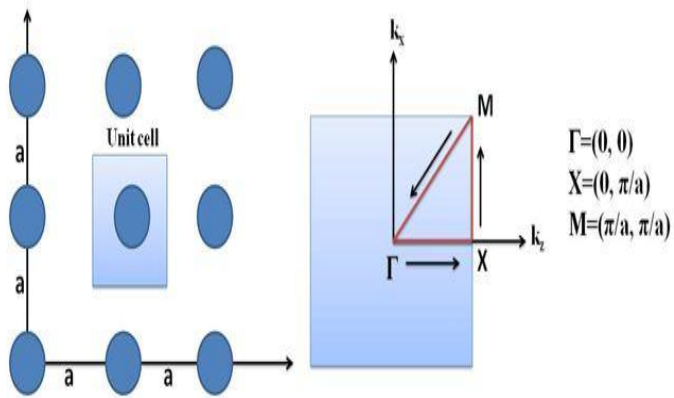

**Figure 2.**

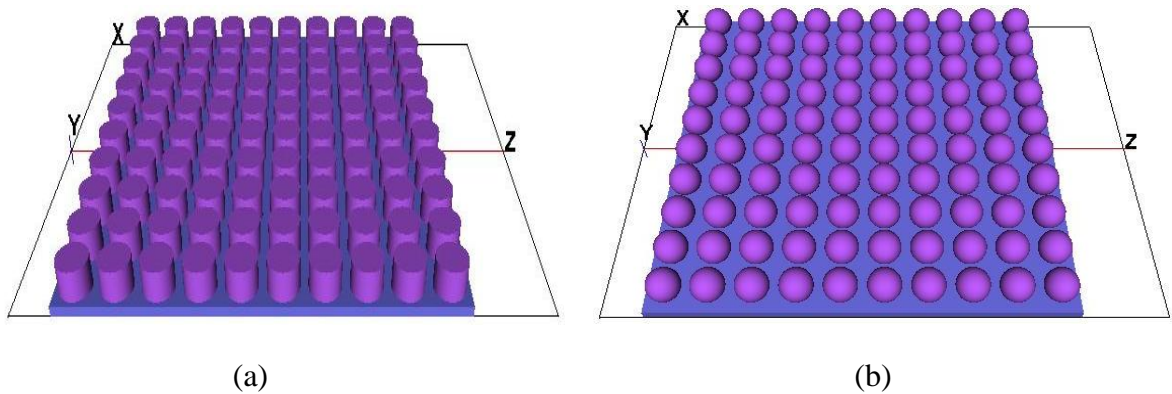

(a)  (b)

**Figure 3.**

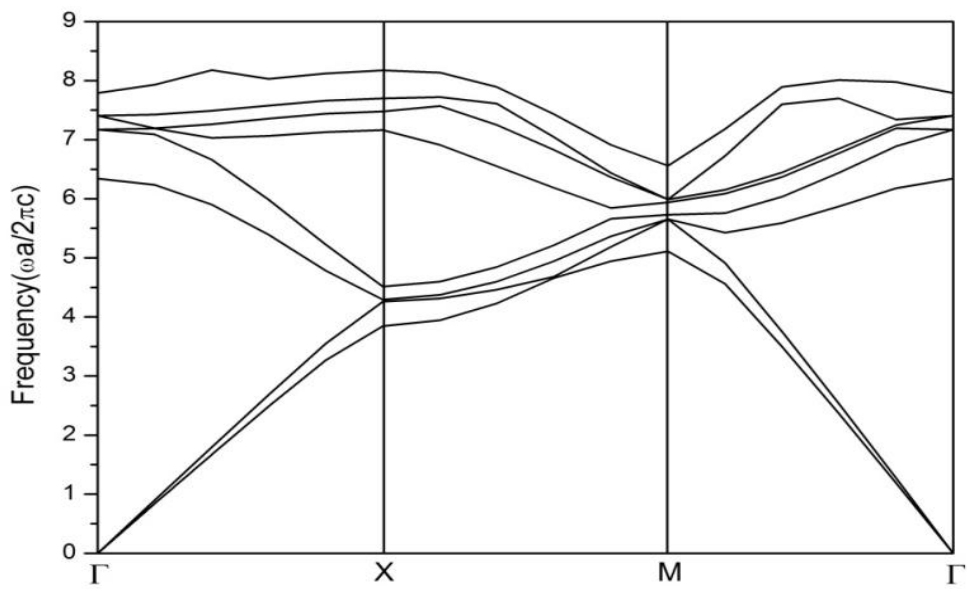

(a)

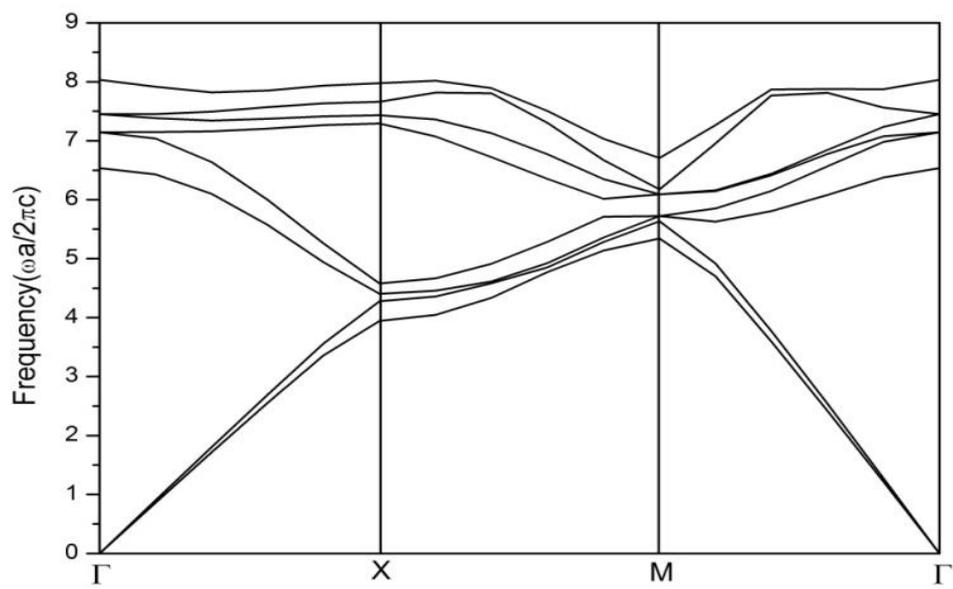

(b)

**Figure 4.**

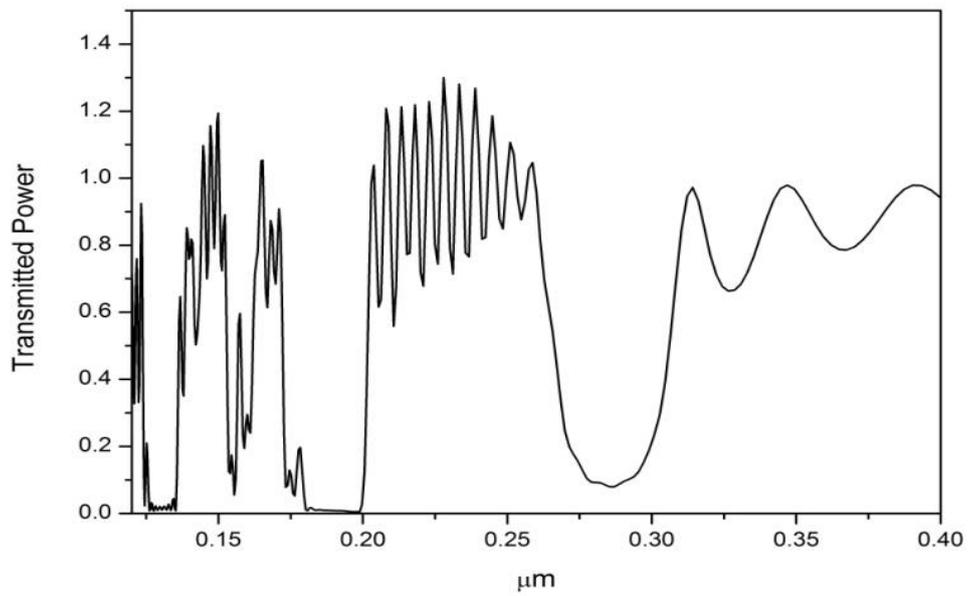

(a)

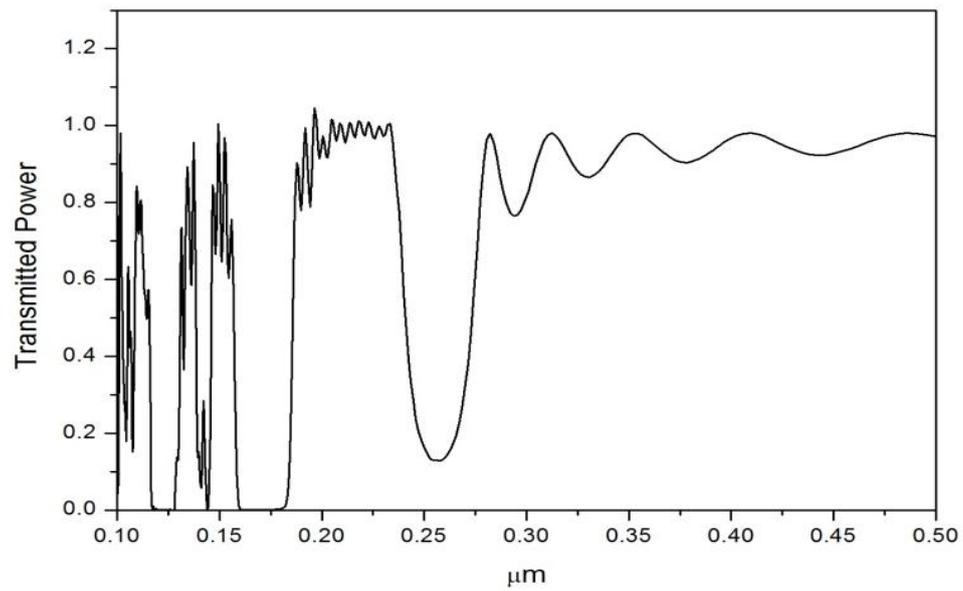

(b)